\documentclass[fleqn,usenatbib,letters]{mnras}

\usepackage{newtxtext,newtxmath}
\usepackage[T1]{fontenc}
\usepackage{ae,aecompl}
\usepackage{xcolor}

\usepackage{graphicx}	% Including figure files
\usepackage{amsmath}	% Advanced maths commands
\usepackage{amssymb}	% Extra maths symbols
\title[Building a strong-lensing cluster watch-list]{On building a cluster watch-list for identifying strongly lensed supernovae, gravitational waves and kilonovae}

\author[D. Ryczanowski et al.]
       {Dan Ryczanowski,$\!^1$\thanks{E-mail: danr@star.sr.bham.ac.uk}
        Graham P.\ Smith,$\!^1$
        Matteo Bianconi,$\!^1$
        Richard Massey,$\!^2$\and
        Andrew Robertson,$\!^2$
        Mathilde Jauzac $\!^{2,3,4}$\\\\
$^{1}$ School of Physics and Astronomy, University of Birmingham, Birmingham, B15 2TT, UK\\
$^{2}$ Centre for Extragalactic Astronomy, Durham University, South Road, Durham DH1 3LE, UK\\
$^{3}$ Institute for Computational Cosmology, Durham University, South Road, Durham DH1 3LE, UK\\
$^{4}$ Astrophysics and Cosmology Research Unit, School of Mathematical Sciences, University of KwaZulu-Natal, Durban 4041, South Africa\\
}

\date{Accepted XXX. Received YYY; in original form ZZZ}

\pubyear{2020}

\begin{document}
\label{firstpage}
\pagerange{\pageref{firstpage}--\pageref{lastpage}}
\maketitle

\newcommand\ls{\ensuremath{\hbox{\rlap{\hbox{\lower4pt\hbox{$\sim$}}}\hbox{$<$}}}}
\newcommand\gs{\ensuremath{\hbox{\rlap{\hbox{\lower4pt\hbox{$\sim$}}}\hbox{$>$}}}}
\defcitealias{NFW1997}{1997}
% Abstract of the paper

\begin{abstract}
Motivated by discovering strongly-lensed supernovae, gravitational waves, and kilonovae in the 2020s, we investigate whether to build a watch-list of clusters based on observed cluster properties (i.e. lens-plane selection) or on the detectability of strongly-lensed background galaxies (i.e. source-plane selection). First, we estimate the fraction of high-redshift transient progenitors that reside in galaxies that are themselves too faint to be detected as being strongly-lensed.  We find $\sim15-50$ per cent of transient progenitors reside in $z = 1-2$ galaxies too faint to be detected in surveys that reach ${\rm AB}\simeq23$, such as the Dark Energy Survey.  This falls to $\ls10$ per cent at depths that will be probed by early data releases of LSST (${\rm AB}\simeq25$).  Second, we estimate a conservative lower limit on the fraction of strong lensing clusters that will be missed by magnitude limited searches for multiply-imaged galaxies and giant arcs due to the faintness of such images. We find that DES-like surveys will miss $\sim75$ per cent of $10^{15}$M$_\odot$ strong lensing clusters, rising to $\sim100$ per cent of $10^{14}$M$_\odot$ clusters. Deeper surveys, such as LSST, will miss $\sim40$ per cent at $10^{15}$M$_\odot$, and $\sim95$ per cent at $10^{14}$M$_\odot$. Our results motivate building a cluster watch-list for strongly-lensed transients that includes those found by lens-plane selection.
\end{abstract}

% Select between one and six entries from the list of approved keywords.
% Don't make up new ones.

%%%%%%%%%%%%%%%%%%%%%%%%%%%%%%%%%%%%%%%%%%%%%%%%%%

%%%%%%%%%%%%%%%%% BODY OF PAPER %%%%%%%%%%%%%%%%%%

\section{Introduction}\label{sec:intro}

The discovery of Supernova Refsdal marked the dawn of observational studies of strongly-lensed transients \citep{KellyLSNe}. A new generation of wide-field optical surveys (e.g. ZTF, \citealt{ZTF}, GOTO, \citealt{GOTO}, LSST \footnote{We refer to the Vera Rubin Observatory as LSST throughout this paper, as the change of name occurred during the time of writing.}, \citealt{LSST}) are now poised to discover large samples of strongly-lensed supernovae (SNe) in the 2020s \citep{GoldsteinLSNeRates,Goldstein2017,Oguri2010}. The prospects for discovering strongly-lensed gravitational waves (GWs) and their electromagnetic counterparts has also been discussed recently, following the early detections of GWs \citep{smith18mnras,smith19wp,smith19obs,NgLensedRates,LiLensedGWRates}.  There is no clear cut evidence that any of the GW events detected thus far are strongly lensed \citep{Hannuksela2019,Singer2019}, due in part to the challenge of detecting a strongly-lensed electromagnetic counterpart and thus localization of a candidate lensed GW to a lens. 

Search strategies for lensed SNe typically involve cross-matching a list of newly discovered transients with a watch-list of luminous red galaxies \citep[e.g.][]{GoldsteinLSNeRates}. Concentrating on individual galaxy lenses is partly driven by the scientific motivation of measuring the Hubble parameter with lensed SNe, which benefits from lower systematic uncertainties arising from simpler lens mass distributions \citep{Bonvin2017,Suyu2017}.  However, strongly lensed SNe and GWs/Kilonovae (KNe) are expected to be dominated by high magnification events, i.e.\ those with lens magnification of $|\mu|>10$, until at least late 2022 when LSST survey operations begin \citep{smith18mnras,GoldsteinLSNeRates}.
Calculations using hydrodynamical simulations were carried out by \citet{RobertsonODSL} which suggest the mass distribution of high-magnification optical depths for lensed point sources is flat over the range $10^{12}$--$10^{14}$M$_{\odot}$, highlighting the prevalence of cluster scale lenses which have typical mass $M_{200} \simeq 10^{14}$M$_{\odot}$. In addition, the cases of high-magnification lensing of point sources detected thus far are indeed dominated by galaxy clusters \citep[e.g.][]{Sharon2005,Oguri2013,KellyLSNe,Sharon2017}. This motivates constructing cluster-based watch-lists for use along with wide-field surveys to detect strongly lensed transients. The search strategy utilising such a watch-list relies on the detections of transient events from wide-field surveys such as ZTF and LSST. Detections found near to a cluster included in a watch-list can then be flagged as candidate strongly lensed transients for follow-up observations, which will confirm if they are indeed lensed. Transient searches are a core component of such surveys, hence the focus of this paper on how to prepare for finding transients lensed by clusters in their alert streams.

The detectability of lensed SNe/KNe within the LSST era has been studied by several authors recently. Based on simulations, \citet{GoldsteinLSNeRates} predict that many hundreds of strongly lensed supernovae will be detected by LSST per year. Whilst these simulations were carried out with galaxy lenses, the results of \citet{RobertsonODSL} indicate that clusters should lens a comparable amount. KN counterparts to lensed GWs are predicted to be detectable with dedicated target of opportunity observations with LSST in red optical bands, specifically the $z$-band, within a few nights of GW detection. A typical source will be located at $z\simeq1-2$ and magnified by $|\mu|\simeq100$ \citep{smith19obs,Smith19GCN}. Transient point sources close to the tangential critical curves of strong-lensing galaxy clusters have also been shown to be recoverable close to the nominal $5\sigma$ detection limit of LSST-like data \citep{smith19wp}. The prospects for detecting lensed SNe and KNe in crowded cluster cores, based on wide-field survey observations and a watch-list of strong-lensing clusters therefore appear promising.

A key question is: how should clusters be selected in order to construct such a watch-list? Strong-lensing clusters can be chosen by searching for visible bright giant arcs and/or multiple images of distant background galaxies \citep[e.g.][]{MarshallSpaceWarps,LenzenArcFinding} -- we refer to this as source-plane selection. Alternatively, they can be chosen based on the inferred projected mass density of cluster cores \citep[e.g.][]{Wong13,EasyCritics} -- a lens-plane selection method. In this letter, we investigate whether relying solely on ``traditional'' source-plane selection is sufficient for creating a watch-list for the discovery of lensed transients. We concentrate on answering two main questions. Firstly, what fraction of the progenitors of GW events/SNe can be found within galaxies whose apparent magnitudes are fainter than the limits of optical wide-field surveys? This is analogous to estimating the fraction of such transients that appear to lack a host galaxy (hereafter, ``hostless'' lensed transients). A hostless lensed transient (whether GW or electromagnetic radiation) may be the first detectable source lensed by a particular lens, in which case source-plane selection alone could not have identified this lens prior to the transient event. Thus a search using a source-plane selected watch-list would miss all such events. Secondly, we ask what fraction of clusters that are capable of strongly-lensing a transient are identifiable as strong lenses by searching for lensed images in wide-field magnitude-limited surveys, i.e. what fraction are identifiable by source-plane selection?
We investigate these questions in Sections \ref{sec:hostless} and \ref{sec:clustersInSurveys} respectively, before summarising our main results and discussing their implications for strongly-lensed transient detection in Section \ref{sec:summary}. All magnitudes quoted are in the AB system, and we assume a flat cosmology consistent with the recent Planck data: $H_{0}$ = 67.8 kms$^{-1}$Mpc$^{-1}$, $\Omega_{M}$ = 0.308 \citep{PLANCK15}. 

\section{Hostless strongly lensed transients}
\label{sec:hostless}
We first consider the first question posed in the introduction, in which transients occur in galaxies that are strongly-lensed and yet not magnified enough to be detected in wide-field photometric surveys.
In this situation, a watch-list of source-plane selected lenses will not include the clusters responsible for lensing these galaxies, and the lensed transients would appear to be hostless.
To quantify how common this scenario is, we consider the fraction of the progenitors of GW events and SNe that reside in strongly lensed galaxies that are too faint to detect in wide-field strong lens searches -- i.e. the fraction of progenitors that will evolve into apparently hostless lensed transients, $f_{\rm hostless}$.
Specifically, we calculate:
\begin{equation}
    \label{eqn:fhostless}
    f_{\rm hostless}=\frac{\int^{L_{\textnormal{lim}}}_{0}L\,\phi(L)\,dL}{\int^{\infty}_{0}L\,\phi(L)\,dL},
\end{equation}
where $\phi(L)$ is the galaxy luminosity function, which is weighted in Equation \ref{eqn:fhostless} by the luminosity $L$, and $L_{\rm lim}$ is a nominal limit below which lensed galaxies are not detectable.  

This formulation is based on the assumption that GWs and SNe form from stellar remnants, and that the stellar population is traced by the stellar light in galaxies.
We base $\phi(L)$ on our Schechter function \citep{SCHECHTER} fits to galaxy number counts from the COSMOS $i$-band selected photometric redshift catalogue \citep{COSMOS}. The fits are performed for galaxies within particular redshift bins, which are described in Section \ref{sec:numDensity}.

It is important to note that the integrals in Equation~\ref{eqn:fhostless} converge for all of our derived values of the Schechter function faint-end slope parameter, as convergence occurs when $\alpha > -2$.  We adopt $z=1$ and $z=2$ as two representative redshifts as they are typical of the redshifts at which strongly lensed transients have and will be detected.  A single representative value of the faint-end slope ($\alpha=-1.2$) is applied to both redshifts, rather than their individual fit values. Our results are insensitive to this choice, and is within the uncertainties from the COSMOS analysis.

\begin{figure}
 \hspace{-3mm}
 \includegraphics[width=1.1\columnwidth]{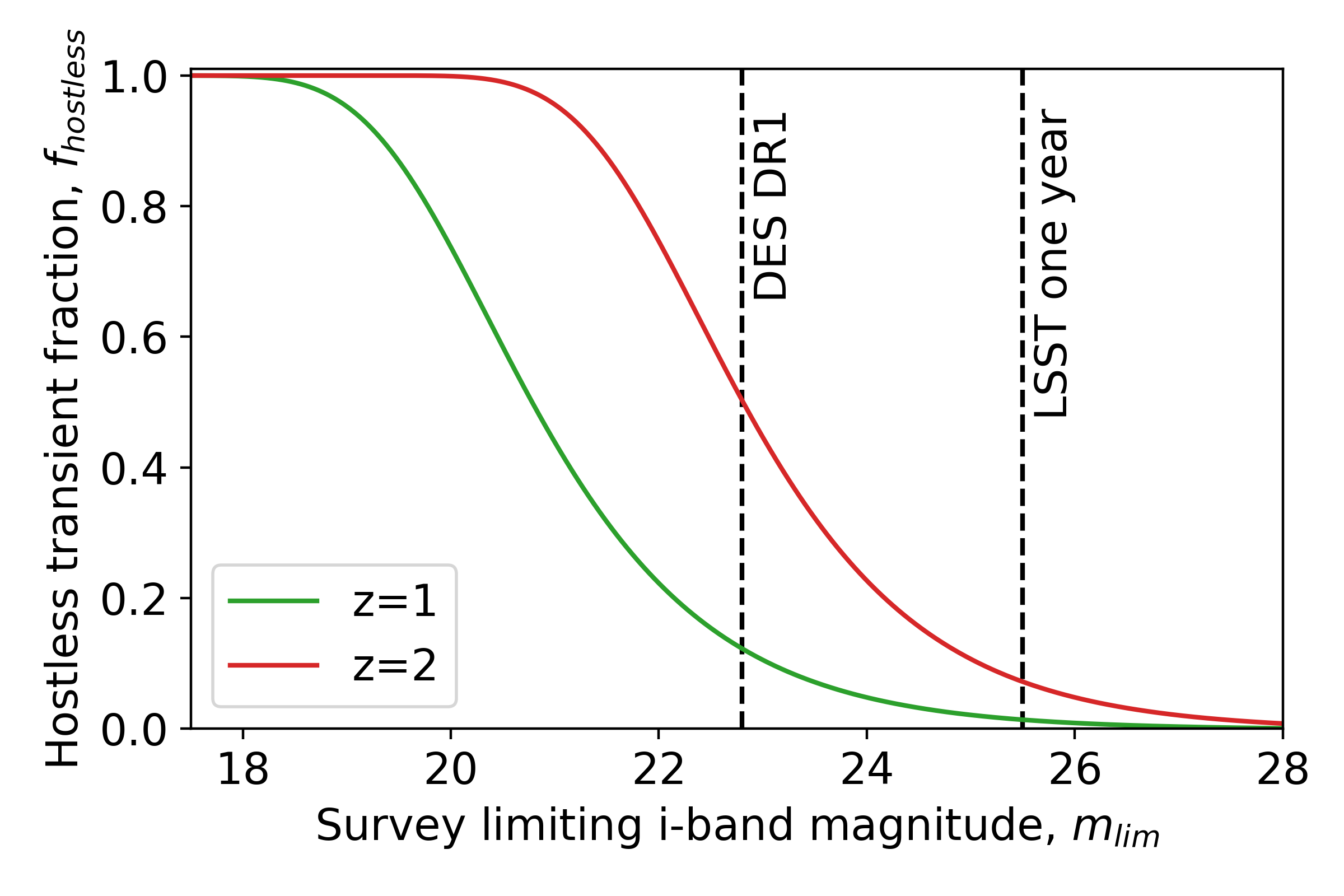}
 \caption{Lower limits on the fraction of strongly-lensed SNe and GWs/KNe that will appear to be hostless. Such transients occur in strongly-lensed galaxies that are too faint to be detected in magnitude-limited searches for strong-lensing clusters based on the detection of multiple images and arcs.  These hostless lensed transients would not initially be identified as lensed, based on strong-lensing cluster watch-lists constructed solely from source-plane searches.  The curves are plotted as a function of the depth to which strongly-lensed galaxies are identified in wide-field survey data.  Prior to the release of data from the first year of LSST \citep[$ i_{\rm lim}\simeq25.5 $][]{LSST}, $\sim15-50$ per cent of the strongly-lensed transients are expected to be hostless, i.e. in lensed $z=1-2$ galaxies beyond the sensitivity limit of precursor surveys such as DES DR1 \citep[$i_{\rm lim}\simeq23.5$][]{DES-DR1}.}
 \label{fig:luminosityDensity}
\end{figure}

\autoref{fig:luminosityDensity} shows $f_{\rm hostless}$ as a function of the limiting $i$-band magnitude of observational surveys.  We relate this limiting magnitude to $L_{\rm lim}$ in Equation~\ref{eqn:fhostless}, by assuming that all images of galaxies that are strongly-lensed by clusters are magnified by $|\mu|=10$.
The plot shows that a significant fraction of transient progenitors, $f_{\rm hostless}\simeq0.15-0.5$, are housed within $z\simeq1-2$ galaxies that are fainter than the detection limits of current wide-field surveys, such as the Dark Energy Survey \citep[DES, ][]{DES-DR1}. This indicates that with present surveys, a significant fraction of lensed transients will be located in lensed galaxies that are not identifiable in magnitude-limited surveys, and thus if the lensed transients are detected they will appear to be hostless.  The situation is less severe for sensitivities comparable with that of LSST after one year of observing, with $f_{\rm hostless} \ls 0.1$.  However, this is little comfort for efforts to detect strongly lensed transients within the first year or two of LSST observations, because cluster watch-lists for these early years of LSST will be constructed from pre-LSST data. It is also important to stress that our estimates of $f_{\rm hostless}$ are lower limits, because they assume that all strongly-lensed galaxies brighter than $i_{\rm lim}$ are detectable -- i.e. source-plane lens detection methods are perfect.

In summary, our estimates of $f_{\rm hostless}$ motivate consideration of lens-plane search methods for finding strong lensing clusters, as they suggest current watch-lists built from purely source-plane search methods may not contain all of the clusters responsible for lensing a non-negligible number of hostless transients.

\section{Strong-lensing clusters in magnitude limited surveys}
\label{sec:clustersInSurveys}

To answer the second question posed in the introduction, we develop a model to estimate the fraction of strong lensing clusters that will be unidentifiable as lenses in magnitude-limited searches for multiply-imaged galaxies or bright arcs. In other words, we estimate the fraction of clusters that would require lens-plane methods to identify, assuming we cannot increase the sensitivity of source-plane searches.

The model utilises Poisson statistics to determine the probability of finding no detectable strongly lensed galaxies behind a cluster lens, based on galaxy number densities from the COSMOS survey. The absence of any detectable galaxies within a lens' so-called strong lensing cross-section implies that no observable bright arcs or multiple images will be seen, and so would not be identifiable as a strong lens in a magnitude-limited source-plane search.  Our estimates of the fraction of unidentifiable strong-lensing clusters are lower limits for the same reason as discussed in Section~\ref{sec:hostless}, i.e.\ that any inefficiency of methods used to search for multiply-imaged galaxies will act to boost the fraction of strong lensing clusters that are missed.

\subsection{Statistical model}
\label{subsec:statModel}

We estimate $f_0$, the fraction of unidentifiable strong lensing clusters of mass $M_{\rm 200}$, from the probability that the number of strongly-lensed galaxies behind a cluster that are brighter than the photometric depth of the survey, $m_{\rm limit}$, is zero:
\begin{equation}
    f_0(M_{\rm 200},m_{\rm limit})=P(0|M_{\rm 200},m_{\rm limit})=\prod_i p_i(0|M_{\rm 200},m_{\rm limit}),
\end{equation}
where the index $i$ runs over a series of background redshift bins.
The Poisson probability of having no detectable sources in the $i$th redshift bin within the strong lensing cross-section of a cluster is given by:

\begin{equation}
    p_i(0|M_{\rm 200},m_{\rm limit})=\exp[-N_i(m_{\rm limit})\,\sigma_i(M_{\rm 200})],
    \label{eqn:prob}
\end{equation}
where $N_i$ is the mean number density of detectable galaxies and $\sigma_i$ is the strong lensing cross-section of the foreground lens. It should be noted that $\sigma_i$ takes different values depending on the redshift of the sources because it depends explicitly on the geometry of the lensing system (Equations \ref{eqn:sisCrossSec} \& \ref{eqn:einsteinradius}).
The overall fraction of unidentifiable strong lensing clusters is therefore given by:
\begin{equation}
    f_0=\exp\left(-\sum_i N_i\,\sigma_i\right).
    \label{eqn:emptyCrossSec}
\end{equation}
Equations~\ref{eqn:prob}~\&~\ref{eqn:emptyCrossSec} assume that galaxies are randomly distributed on the sky, and therefore ignore galaxy clustering. Clustering concentrates some of the galaxy population into particular regions of space near to each other, meaning a randomly-positioned aperture of fixed size will find zero galaxies within it more frequently than using the above Poisson-based calculations. Using Equation~\ref{eqn:emptyCrossSec} will therefore result in a conservative lower limit on $f_0$.

\subsection{Strong lensing cross-section}
\label{subsec:SL_cross-sec}
Our model for the strong lensing cross-section, $\sigma$, of a lens relies on an analytic description of the typical mass distribution in galaxy clusters on the scale of the Einstein radius -- i.e. a few tens of arcseconds.  The singular isothermal sphere (SIS) model is commonly used to quantify the density profile in studies that predict the lensing properties of massive galaxies \citep[e.g.][]{Gavazzi2007}.  This model has a projected density profile of:
\begin{equation}
    \Sigma(R) = \frac{\sigma_{v}^{2}}{2GR},
    \label{eqn:SISdensity}
\end{equation}
where $G$ is the gravitational constant, $R$ is the projected radial distance from the centre of the lens and $\sigma_{v}$ is the velocity dispersion. However, the projected density profile of clusters in the strong-lensing regime is typically shallower than the $R^{-1}$ isothermal profile.  For example, the slope of the projected density profile of strong-lensing clusters in the Local Cluster Substructure Survey (LoCuSS) sample span exponents in the range $-0.11$ to $-0.87$ at the respective Einstein radii of the clusters for sources at $z_S=2$, with a median of $-0.57$ (\citealt{RichardClusterModel}; see also Figure 3 of \citealt{Umetsu2016}).

A common alternative to the SIS model is the Navarro, Frenk and White (NFW, \citetalias{NFW1997}) model. Whilst this model describes the observed density profile of clusters well on large scales \citep[e.g.][]{OkabeSmithMassMeasures}, it is too shallow within the Einstein radius of clusters. For example, the NFW models described in \citet{OkabeSmithMassMeasures} that fit the weak-lensing constraints on the clusters in the \citet{RichardClusterModel} strong-lensing sample have exponents in the range $-0.22$ to $-0.41$ at the respective Einstein radii of the clusters for a source at $z_S=2$.  Therefore neither the SIS nor the NFW model alone provides a completely faithful description of the azimuthally averaged properties of strong-lensing clusters clusters at their Einstein radii -- the SIS model is too steep and the NFW model is too shallow. We adopt the SIS model because its slightly steeper slope will boost the strong-lensing cross section relative to reality and thus ensure that our end values of $f_0$ (Eq. \ref{eqn:emptyCrossSec}) are conservative.  After describing our definition of cross section below, we will return to this point at the end of this Section.

For simplicity, we consider a galaxy to be strongly lensed if it experiences a gravitational magnification of $|\mu|\ge10$. We also assume that all strongly lensed galaxies suffer the same magnification $|\mu| = 10$ -- the smallest value we associate with strong lensing -- and ignore the effect of larger magnifications. This is justified, as the quantity of strong lensing regions with the capability to produce a magnification greater than some value $|\mu|$ falls off as $\mu^{-2}$ \citep{Blandford1986}, and so the population of strong lensing lines of sight is dominated by those of lower magnification. Defining strong lensing in this way allows us to quantify the strong lensing cross-section by finding the region within which the magnification exceeds a particular value of $|\mu|$. Following the equations describing the SIS model in e.g. \citet{DodelsonBook}, this condition is found to be satisfied by at least one image when the source is within the region:
\begin{equation}
    \label{eqn:SLregion}
    \beta < \frac{\theta_{\rm E}}{|\mu|-1},
\end{equation} 
where $\beta$ is the angular separation from the centre of the lens in the source plane and $\theta_{\rm E}$ is the Einstein radius of the lens. Using Equation \ref{eqn:SLregion}, the strong lensing cross-section can be found by taking advantage of the symmetry of the SIS model and using our $|\mu|=10$ definition of strong lensing:
\begin{equation}
    \label{eqn:sisCrossSec}
    \sigma=\pi\beta^{2}=\frac{\pi\theta_{\rm E}^{2}}{81},
\end{equation}
which depends only on the Einstein radius. Within the SIS model, the Einstein radius can be expressed in terms of the lens mass as:
\begin{equation}
    \theta_E = \frac{2\pi}{c^{2}}\left(\frac{D_{\rm LS}}{D_{\rm S}}\right)(10\,G\,H_{\rm L}\,M_{200})^{\frac{2}{3}},
    \label{eqn:einsteinradius}
\end{equation}
where $H_{\rm L}$ is the Hubble parameter at the lens redshift, and $D_{\rm LS}$ and $D_{\rm S}$ are the angular diameter distances between the lens and the source, and the observer and the source respectively. $M_{200}$ is defined as the mass within a radius $r_{200}$, within which the mean density of the cluster is 200 times the critical density of the universe. 
We adopt a single lens plane at a redshift of $z_{\rm L}=0.25$, which fits with the binning procedure used in Section~\ref{sec:numDensity}, and is close to the median value from the sample of known strong lensing clusters discussed by \citet{smith18mnras}. Following these specifications, the strong lensing cross-section within each redshift bin can be determined as a function of halo mass.

The range of density profile slopes for known strong-lensing clusters noted above implies considerable scatter in their cross-sections. This could be due to a variety of phenomena such as substructures within cluster cores that are associated with cluster-cluster and cluster-group mergers \citep[e.g.][]{Smith05,Bradac08,Limousin12,Jauzac19}, cluster halo triaxiality \citep[e.g.][e.g.]{Oguri05,Corless09,SerenoZitrin12,Sereno13}, or line-of-sight structure \citep{Wambsganss2005, Bayliss2014, DAloisio2014}. We investigate this scatter further in the context of our definition of the lensing cross-section, in part as a cross-check on our assertion that adopting the SIS model is conservative.  We consider the fourteen X-ray selected clusters in common between the \citet{RichardClusterModel} strong-lensing analysis and \citet{OkabeSmithMassMeasures} weak-lensing analysis.  We used the models of \citeauthor[][]{RichardClusterModel} to compute the total source-plane solid angle that is magnified by $|\mu|\geq10$, and compared these values with the equivalent SIS cross-sections that are calculated using Equation 6 and $M_{200}$ for each cluster from \citeauthor{OkabeSmithMassMeasures}. The distribution of the ratio of SIS to LoCuSS cross-section is approximately log-normal and is centred at $\ln(\sigma_{SIS}/\sigma_{LoCuSS}) = 0.23$, implying that the SIS model typically over-estimates the lens cross-sections by 26 per cent -- i.e. in qualitative agreement with our expectations.  The standard deviation of the distribution implies a factor of $\sim3.7$ scatter around the central value of 0.23, which reflects the structural diversity of strong-lensing cluster cores.  Therefore, when discussing cluster mass in later Sections, it should be taken to mean ``clusters that have cross-sections comparable with the typical cluster of that mass''.  The scatter therefore raises important questions about how to approach the lens-plane selection of strong lensing clusters, and does not alter the broad conclusions of this letter.  We will investigate methods to lens-plane select strong lensing clusters in future work.

A final consideration to be made is of the finite size of galaxies in the source plane. The typical solid angles of optically-selected galaxies from \citet{FergusonGalaxySize} was compared to the SIS cross-section, and it was found that typical $z=2$ galaxies subtend a solid angle comparable to the SIS cross-section of an $M_{200} = 5\times10^{13}$M$_\odot$ cluster. Therefore less massive clusters do not strongly lens a region large enough to enclose an entire typical galaxy. Because of the galaxy number densities (Section \ref{sec:numDensity}, this caveat does not turn out to affect the conclusions of this study. This is because the fraction of clusters that lens one or more galaxies does not become significant until at least $M_{200} \sim 10^{14}$M$_\odot$ (depending on observation depth, see Section \ref{sec:predictions}), at which point the cross-section is much larger.

\subsection{Galaxy number density}
\label{sec:numDensity}
Galaxy number densities, $N_{i}$, were estimated using number counts from the COSMOS catalogue of \cite{COSMOS}. Galaxies are binned so that the observable number density can be quantified as a function of redshift. The catalogue was split up into eleven redshift bins in the range $0.25<z<5.75$, each with a constant width of $0.5$. This scheme was chosen because it spans the majority of the catalogue (which includes objects up to redshift $z=6$) and allowed for the lens to reside at a redshift $z_{\rm L}=0.25$, a value consistent with the current known population of strong lensing clusters \citep{smith18mnras}. Ultimately, the final three bins centred at $z = 4.5$, $5.0$ and $5.5$ were excluded from the analysis as they contained very few galaxies with well-constrained redshifts.

The photometric depth of the catalogue from \citeauthor{COSMOS}, $i=25$, is well matched to the first data release (DR1) of DES, which reaches approximately this magnitude after taking into account a lens magnification of $|\mu| = 10$. However, data from the final release of DES, and other surveys including upcoming data from LSST will reach depths up to $i\simeq28$ after including this magnification, and hence will probe intrinsically fainter galaxies beyond those of the COSMOS catalogue. Therefore, we fit a Schechter function \citep{SCHECHTER} to the number counts in each redshift bin, allowing the COSMOS number counts to be extrapolated when considering deeper surveys.

The dominant source of uncertainty in this extrapolation is the so-called ``catastrophic failure'' rate of the photometric redshifts. The catastrophic failure rate quantifies how often photometric redshift measurements differ significantly from reliable spectroscopic redshift measurements, as defined in \citet{COSMOS}. If the rate is high, then many galaxies will be placed in the wrong redshift bins, skewing the number densities and hence affecting the curve fitting and extrapolation. The failure rates provided in the catalogue paper \citep{COSMOS} vary with apparent magnitude, but are of order 15 to 20 per cent for the faintest sources (which \citeauthor{COSMOS} categorise as $i>23$). We quantify the overall effect of catastrophic failures on the end result by considering a worst-case scenario based on the quoted failure rates, and in this scenario determine how many galaxies would appear in the wrong redshift bins. Then, correct for this and re-do the calculation based on the new ``corrected'' number counts. The true probability curve will then lie somewhere between the original and corrected result, providing a range of uncertainty on the final result. Doing so provides an error on the final result of $<2$ per cent. 

\subsection{Predictions of the model}
\label{sec:predictions}
\begin{figure}
 \includegraphics[width=\columnwidth]{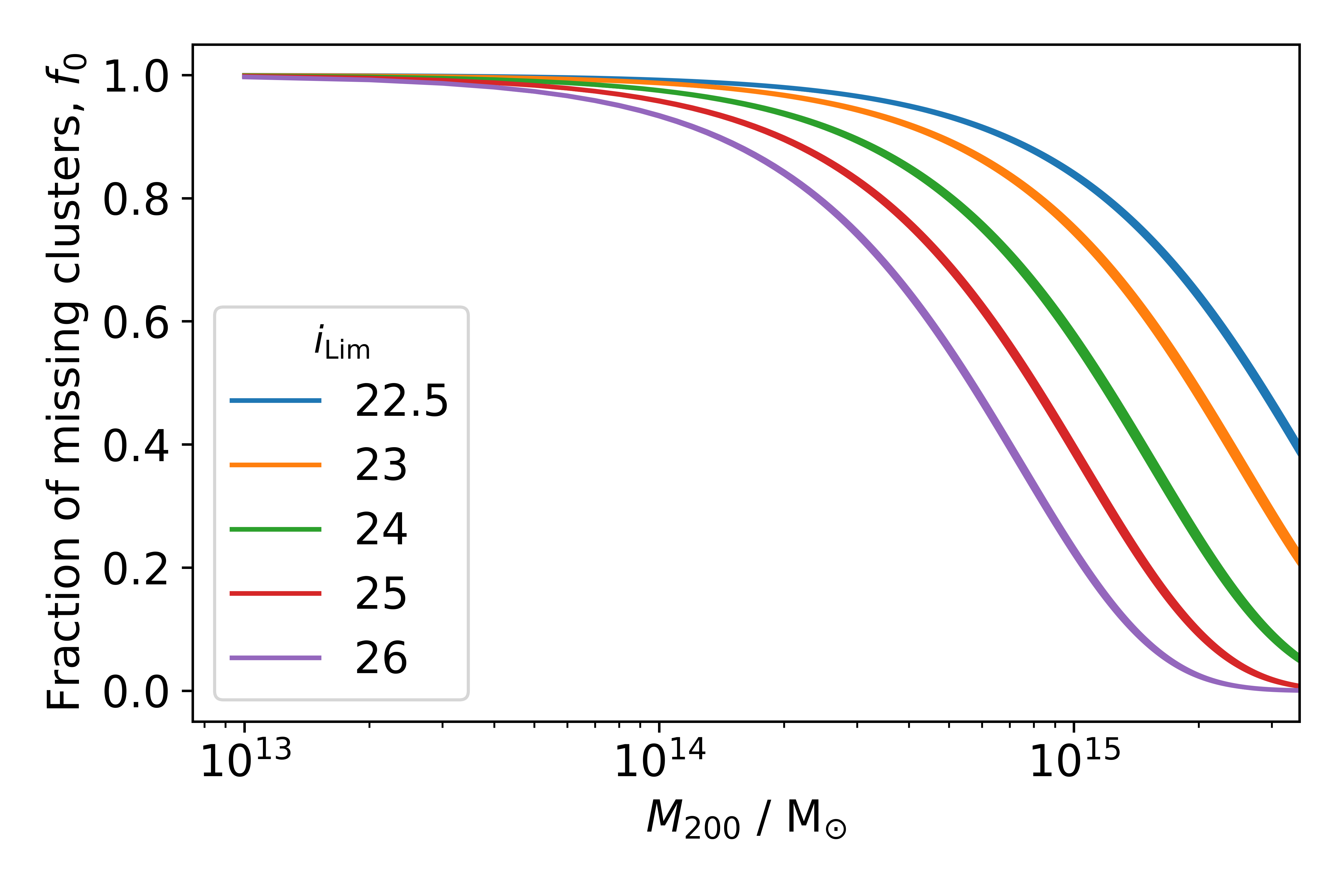}
 \caption{The fraction of strong lensing clusters that are unidentifiable as strong lenses in magnitude-limited surveys, as a function of halo mass, $M_{200}$, and survey depth, $i_{\rm lim}$. The coloured lines represent various magnitude depths similar to those of recent and upcoming surveys. The thickness of each line corresponds to the uncertainty due to photometric redshift failures.
 DES-like surveys will miss $\sim75$ per cent of $10^{15}$M$_\odot$ strong-lensing clusters, rising to $\sim100$ per cent at $\sim10^{14}$M$_\odot$.  Deeper surveys such as LSST will miss $\sim40$ and $\sim95$ per cent at these masses.}
 \label{fig:moneyplot}
\end{figure}

\autoref{fig:moneyplot} shows our estimated lower limit on the fraction of strong-lensing clusters that will be unidentifiable by wide-field surveys as a function of the lens halo mass and survey magnitude depth, as calculated by the model described in Section~\ref{sec:clustersInSurveys}. The fractions are calculated for various survey depths that span ongoing and upcoming surveys. The uncertainties in each curve, represented by their width, are due to limitations of the photometric redshifts in the COSMOS data set, as outlined in Section \ref{sec:numDensity}. This plot shows that for the most massive strong lensing galaxy clusters, with masses of $\sim10^{15}$M$_\odot$, about 75 per cent would not be identifiable as strong lenses by a survey similar to DES DR1, i.e.\ $i_{\rm lim}\simeq23.5$ \citep{DES-DR1}. In deeper surveys, such as the first year of LSST observations ($i\simeq25.5$), only around 40 per cent of $10^{15}$M$_\odot$ will not be identifiable as strong lenses. Lenses of lower mass are even less likely to be identified. The model predicts that even with deeper surveys, no more than 5 per cent of the more common $10^{14}$M$_{\odot}$ strong lensing capable clusters will be identifiable by searching for multiply imaged galaxies or bright arcs. We therefore conclude that the level of incompleteness of source-plane searches will be particularly severe for strong lensing clusters with typical masses close to the knee of the halo mass function, independent of whether deep future LSST survey data are available.

\section{Summary and Implications for Strongly-lensed Transients}
\label{sec:summary}
Observational results \citep{Sharon2005,Oguri2013,KellyLSNe,Sharon2017,Kelly2018NatAst,Rodney2018NatAst} and theoretical predictions \citep{Hilbert2008,RobertsonODSL} all point to galaxy clusters making a significant contribution to the high-magnification strong lensing optical depth of point sources.  This implies that some of the strongly-lensed transients discoverable by ongoing and future optical surveys (e.g. ZTF, GOTO, LSST) will be strongly-lensed by galaxy clusters.  
So far, searches for transients strongly-lensed (i.e. multiply-imaged) by clusters have employed pointed observations of known strong-lensing clusters -- i.e.\ clusters that are known to be strong lenses because multiply-imaged galaxies have been detected in their central regions \citep[e.g.][]{Goobar2009,KellyLSNe,Rubin2018,smith19obs}, or in other words, clusters selected in the source plane.  In this study we have investigated whether it is beneficial to build a watch-list of strong-lensing clusters for wide-field optical surveys such as LSST, based on only known strong lenses (selected in the source plane) or on a more extensive (likely more complete but less pure) list of strong-lensing capable clusters which includes those selected in the lens plane.

First, in Section \ref{sec:hostless} and Figure~\ref{fig:luminosityDensity}, we derived a lower limit on the fraction of lensed transients whose host galaxies are fainter than the magnitude limit of wide-field photometric surveys, even after taking lens magnification into account. Such lensed transient point sources would appear to be hostless and would be lensed by clusters that are not identifiable as strong lenses by a source-plane search. We predict that the fraction of hostless lensed transients is $f_{\rm hostless}\simeq0.15-0.5$ for cluster watch-lists that are based on source-plane selection in data of depth similar to DES DR1.  This falls to $f_{\rm hostless}\ls0.1$ for source-plane-based watch-lists derived from LSST year one data. Our estimates of $f_{\rm hostless}$ therefore imply that galaxy clusters that are not identifiable as strong lenses in magnitude limited surveys should be included in watch-lists so that the non-negligible number of lensed hostless transients can be identified, especially if the watch-list is based on relatively shallow data before the LSST survey begins.

Second, in Section~\ref{sec:clustersInSurveys} and Figure~\ref{fig:moneyplot}, we predict the fraction of strong lensing capable clusters that are unidentifiable due to an absence of detectable multiply-imaged galaxies.  We predict that the fraction of unidentifiable cluster lenses of mass $M_{200}\simeq10^{15}$M$_\odot$ is $f_0\simeq0.75$ at depths comparable to DES DR1, and $f_0\simeq0.4$ at depths comparable to the first year of LSST observations. For more abundant $10^{14}$M$_\odot$ clusters, we predict that even at LSST one-year depth, the fraction of unidentifiable strong lensing clusters is $f_0\gs0.95$. We emphasize that these predictions are conservative lower limits as we assume no galaxy clustering and a perfectly efficient source-plane cluster strong-lens search algorithm.
Taken together, our results on $f_{\rm hostless}$ and $f_0$ both motivate building cluster watch-lists for strong-lensing transient discovery based on lens-plane selection. In future work we will explore methods for lens-plane selection.

\section{Acknowledgements}
DR (ORCID 0000-0002-4429-3429) acknowledges a PhD studentship from the Science and Technology Facilities Council. GPS (ORCID 0000-0003-4494-8277) and MB (ORCID 0000-0002-0427-5373) acknowledge support from the Science and Technology Facilities Council through grant number ST/S000305/1. AR (ORCID 0000-0002-0086-0524) is supported by the European Research Council (ERCStG-716532-PUNCA) and the STFC (ST/N001494/1). RM (ORCID 0000-0002-6085-3780) acknowledges the support of a Royal Society University Research Fellowship. MJ (ORCID 0000-0003-1974-8732) is supported by the United Kingdom Research and Innovation (UKRI) Future Leaders Fellowship 'Using Cosmic Beasts to uncover the Nature of Dark Matter' [grant number MR/S017216/1]. This project was also supported by the Science and Technology Facilities Council [grant number ST/L00075X/1]. We thank Johan Richard for sharing the lens models from Richard et al. (2010) with us, and acknowledge helpful discussions with Rahul Biswas, Martin Freer, Ariel Goobar, Matteo Maturi, Matt Nicholl, and Evan Ridley.

%%%%%%%%%%%%%%%%%%%%%%%%%%%%%%%%%%%%%%%%%%%%%%%%%%

%%%%%%%%%%%%%%%%%%%% REFERENCES %%%%%%%%%%%%%%%%%%

% The best way to enter references is to use BibTeX:

\bibliographystyle{mnras}
\bibliography{refs}

% Alternatively you could enter them by hand, like this:
% This method is tedious and prone to error if you have lots of references
%\begin{thebibliography}{99}
%\bibitem[\protect\citeauthoryear{Author}{2012}]{Author2012}
%Author A.~N., 2013, Journal of Improbable Astronomy, 1, 1
%\bibitem[\protect\citeauthoryear{Others}{2013}]{Others2013}
%Others S., 2012, Journal of Interesting Stuff, 17, 198
%\end{thebibliography}

%%%%%%%%%%%%%%%%%%%%%%%%%%%%%%%%%%%%%%%%%%%%%%%%%%

%%%%%%%%%%%%%%%%% APPENDICES %%%%%%%%%%%%%%%%%%%%%

%%%%%%%%%%%%%%%%%%%%%%%%%%%%%%%%%%%%%%%%%%%%%%%%%%

% Don't change these lines
\bsp	% typesetting comment
\label{lastpage}
\end{document}